\documentclass[cits]{PoS}

\usepackage{amsmath,amsthm,amsfonts,amssymb}
\usepackage[utf8]{inputenc}
\usepackage{subfig}

\renewcommand{\href}[1]{}
\newcommand{\path}[1]{#1}

\title{Gauge Field Generation on Large-Scale GPU-Enabled Systems}

\ShortTitle{Gauge Field Generation on GPUs}

\author{\speaker{Frank~Winter}\\
         School of Physics and Astronomy,
                    University of Edinburgh,
                    Edinburgh EH9 3JZ, UK \\
        E-mail: \email{frank.winter@ed.ac.uk}}

\abstract{
Over the past years GPUs have been successfully applied to the task of
inverting the fermion matrix in lattice QCD calculations.
Even strong scaling to capability-level supercomputers, corresponding
to O(100) GPUs or more has been achieved.
However strong scaling a whole gauge field generation algorithm to
this regim requires significantly more functionality than just having
the matrix inverter utilizing the GPUs and has not yet been
accomplished.
This contribution extends QDP-JIT, the migration of SciDAC QDP++ to
GPU-enabled parallel systems, to help to strong scale the whole
Hybrid Monte-Carlo to this regime.
Initial results are shown for gauge field generation with Chroma
simulating pure Wilson fermions on OLCF TitanDev.
}

\FullConference{
The 30th International Symposium on Lattice Field Theory, Lattice 2012\\
June 24-29, 2012\\
Cairns, Australia}

\begin{document}

\section{Introduction}

Gauge field generation in lattice QCD is a computationally demanding
task typically requiring capability-level supercomputers. 
The task of inverting the fermion matrix accounts for the vast
majority of floating-point operations in such simulations.
GPUs have been applied with great success to this task and with
suitable algorithms even strong scaling to the capability regime is
possible
\cite{Clark:2009wm, Babich:2010mu, Babich:2011np}. 
However, gauge field generation algorithms like the Hybrid Monte-Carlo 
(HMC) require substantially more functionality than just the matrix
inversions.
They typically require a pseudofermion refresh, momentum updates,
force term calculation, projection of fields, reunitarization of
SU(3) matrices,
a random number generator, exponentiating of SU(3) matrices, and
more.
When leaving those operations not accelerated huge effects due to
Amdahl's law render the whole algorithm (including the fermion matrix)
only weakly scaling to the capability regime.

The Chroma package is a versatile lattice QCD software suite with a
broad area of application \cite{Edwards:2004sx}. 
This includes the post-Monte Carlo analysis phase as well as gauge
field generation. 
Chroma implements its operations using the high-level data-parallel
programming interface QDP++.
However QDP++ was not designed for heterogeneous architectures like
GPUs-enabled systems.

QDP-JIT is an implementation of the QDP++ API especially designed for
GPU-enabled systems and was introduced in earlier work
\cite{Winter:2011dh} (there, however, operating only a single GPU).
QDP++ operations are off-loaded automatically to the accelerators,
Just-In-Time (JIT) compilation using NVIDIA NVCC generates the GPU
kernels.
This work extends QDP-JIT with
\begin{itemize}
\item an automated memory management,
\item multi-GPU (parallel architecture) support,
\item auto-tuning of CUDA kernels
\item support for non-linear algebra function like \texttt{sum}, \texttt{sumMulti}, etc.
\end{itemize}

This provides a 100\% QDP++ implementation for GPU-enabled systems
effectively executing all operations in Chroma on the accelerators.

\subsection{Why JIT Compilation?}

Typically JIT compilers are used in systems executing managed
software, dynamically typed languages and in optimization through
function specialization.
Why would a lattice QCD application need a JIT compiler? 
In the case where function assembly is done with the host compiler
(CPU architecture) using compile-time code transformations,
e.g. expression templates (ET) as found in QDP++, and the target
architecture (GPU) being different from the host architecture dynamic
code generation is required unless the compiler is changed.
The rationale is that no cross-compiler (CPU host, GPU target) exists
that allowed a C++ API to kernel code. 
NVIDIA NVCC comes close but provides only a C interface to accelerated
code.
Neither on the QDP++ side it is desirable to abdicate the use of ETs
due to the optimizations they provide like improved cache locality and a
reduced number of site loops, nor is it anticipated at this stage to
move to a compiler framework.

\section{Dynamic Code Generation}

For the purpose of generating CUDA kernel code QDP-JIT extends the ET
implementation found in QDP++ by a pretty printing functionality that
writes a C code equivalent in prefix notation of the represented
expression to the filesystem. 
Subsequently a system call launches NVIDIA NVCC which generates
position-independent GPU code wrapped into a cubin fat binary.
The Linux linking loader is then used to map the code into the current
process and finally a call to the function is issued.

NVIDIA NVCC is meant for static compilation and has a significant run-time
overhead if used as a JIT compiler. 
Thus kernel generation at runtime is expensive. 
In order to avoid redundant compilation shared objects from prior runs
are kept on the filesystem and pre-loaded upon program startup.

\section{CUDA Thread Parallelization Strategy}

\begin{table}
\begin{center}
\begin{tabular}{ 
    p{0.15\columnwidth}p{0.01\columnwidth}
    p{0.15\columnwidth}p{0.01\columnwidth}
    p{0.15\columnwidth}p{0.01\columnwidth}
    p{0.15\columnwidth} }
Outer & $\times$ & Spin & $\times$ & Color & $\times$ & Reality  \\
\hline
Scalar & & Scalar && Scalar && Real \\
Lattice & & Vector && Vector && Complex \\
        & & Matrix && Matrix &&    \\
        & &        && Seed   &&    \\
\end{tabular}
\end{center}
\caption{
\label{tab:qdp}
QDP++ tensor structure resembles QCD data types.
CUDA thread parallelization is implemented on the outer level
assigning one lattice site to each thread.}
\end{table}

 GPU thread parallelization is implemented on the QDP++ outer level
 (see Tab. \ref{tab:qdp}). Each lattice site is assigned to a
 different CUDA thread. Thus the QDP++ outer level is absent in kernel
 code. All other levels (primitives and reality) are preserved in
 kernel code and inherited from QDP++.

 On NVIDIA Fermi architecture this approach requires the local volume
 to be large enough in order to get a good Streaming Multiprocessor (SM)
 occupancy.

\section{Auto-tuning CUDA kernels}

\begin{figure}[t]
\begin{center}
\includegraphics[width=1.0\columnwidth]{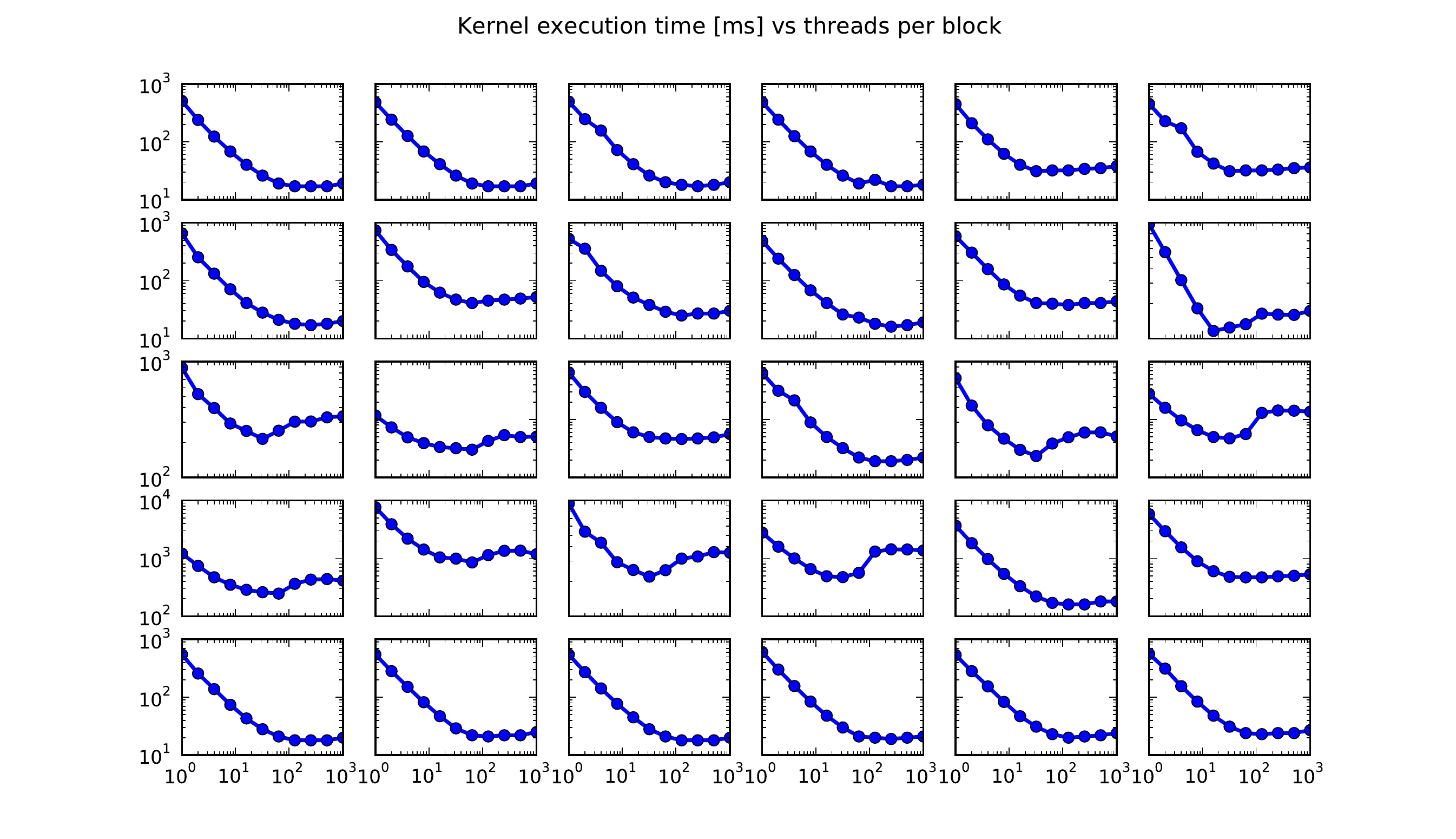}
\end{center}
\vspace*{-3mm}
\caption{\label{fig:tuning}
 Auto tuning of CUDA kernels. Shown is the execution time as function
 of $N_t$ for various kernels for a fixed local volume. The
 performance depends heavily on $N_t$.
}
\end{figure}

In order to improve performance on a per kernel basis QDP-JIT employs
an automatic tuning of the kernels. 
In general CUDA kernel performance depends on: SM occupancy,
arithmetic intensity, and memory access pattern. 
While the latter two at time of kernel launch are fixed SM occupancy
can be tuned with a kernel launch parameter: threads per block $N_t$. 
QDP-JIT executes each kernel and measures the execution time as a
function of $N_t$ and determines the best value and stores it in a
database.
 Fig. \ref{fig:tuning} shows some examples of kernel execution times
 as a function of $N_t$. 

\section{Memory Management}

On-board GPU memory is a limited resource and constrains the
way the memory management should function. 
QDP-JIT implements an automatic memory management which frees the
domain expert from the task of explicitly transferring data between
different memory domains (CPU/GPU), and which allows an unlimited
number of in-scope lattice objects, and minimizes memory transfers
between the host and the accelerator automatically.
The memory management implements a cache with a ``first use'' (GPU or
CPU) allocation strategy. Memory is allocated upon first use instead
of at object creation time and is only allocated in the memory domain
where the object is used.
It also implements a dynamic re-allocation of memory to a different
memory domain. 
The cache uses a least recently used spilling algorithm.

 Since CUDA memory allocation functions are specified as not being fast
 QDP-JIT implements its own allocator similar to a small object
 allocator but with additional cache functionality to ensure true
 $O(1)$ complexity for fast allocation and to avoid memory
 fragmentation.
In QDP-JIT also small objects (like scalars) take part in dynamic
memory management and use a separate allocator operating on a memory
pool registered (page-locked) with the NVIDIA driver to ensure
asynchronous transfers.

\subsection{CUDA Stream API}

\begin{figure}[t]
\begin{center}
\includegraphics[width=0.95\columnwidth]{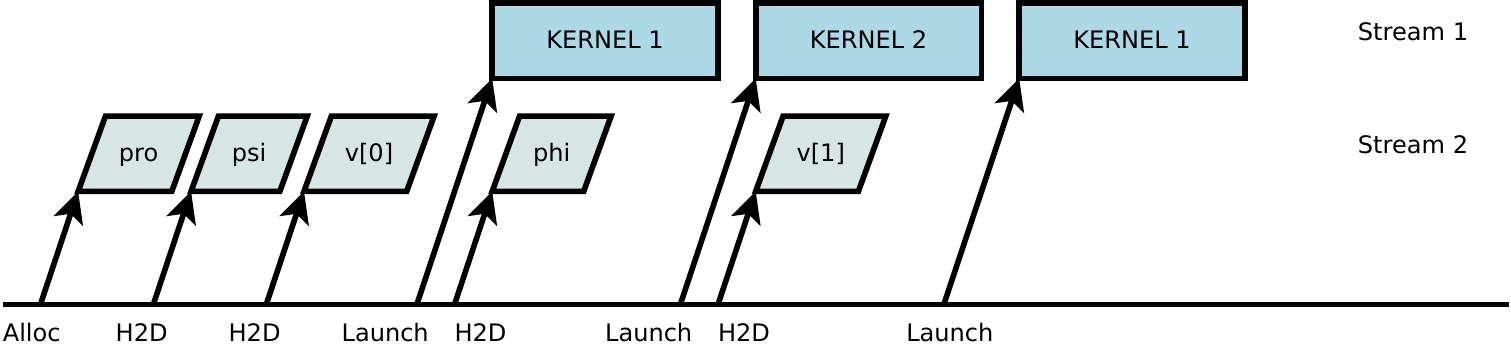}
\end{center}
\vspace*{-3mm}
\caption{\label{fig:stream}
CUDA stream API allows to execute H2D and D2H transfers asynchronously
with host thread and kernel execution. In this example
\texttt{pro,psi,phi,v[0],v[1]} are fields to be copied to the
accelerator.
}
\end{figure}

QDP-JIT uses the CUDA stream API to overlap GPU computation with data
transfers and CPU host thread progress.
Host-to-device (H2D) and device-to-host (D2H) transfers execute
asynchronously with respect to the host thread and (when possible)
overlap with GPU kernel execution, see Fig. \ref{fig:stream}.

\section{Interoperability with QUDA}
    
QDP-JIT was designed to inter-operate with QUDA, a Krylov space solver
package for NVIDIA GPUs 
\cite{Clark:2009wm, Babich:2010mu, Babich:2011np}. 
QUDA's device memory allocation is redirected to the QDP-JIT memory
management. This ensures that both GPU packages (QDP-JIT, QUDA) can be used at the
same time in an application like Chroma: All matrix inversions are
carried out using QUDA whereas QDP-JIT executes all other
operations on the GPU.

\section{Demonstration: HMC running on TitanDev}

\subsection{OLCF Titan}

 The Oak Ridge Leadership Computing Facility (OLCF) has completed the
 upgrade of the Jaguar system to the hybrid-architecture Cray XK7
 system named Titan. The system comprises AMD 16-core Opteron 6274
 processors running at 2.2 GHz. It is accelerated with 18688
 Kepler-based K20s GPUs (1 GPU per 
 node) and currently holds the top position of the Top 500 list, a
 ranking of supercomputers. Earlier in the year 2012
 phase I of this upgrade populated 960 of the Jaguar nodes with Cray
 XK6 nodes with NVIDIA Fermi GPUs. This system was called
 TitanDev. Due to the availability of this system the QDP-JIT
 demonstration was run on TitanDev.

\subsection{Program Setup}

$N_f=2$ flavors of dynamical Wilson fermions were simulated in single
precision on partitions of 16 to 256 Cray XK6 nodes with regular HMC
with Hasenbusch preconditioning \cite{Hasenbusch:2001ne}.
The $32^3\times 96$ lattice configuration to start the HMC was taken
from \cite{Bulava:2009jb}.
Multi-timescale integrators were utilized together with a
chronological inverter with minimal residual extrapolation
\cite{Brower:1995vx}.

\subsection{No JIT on Cray XK6}

The Cray Linux Environment on the compute nodes support the exec
system call which is required for the JIT approach to work.
However the NVIDIA NVCC does not run successfully on the Cray XK6 compute nodes
due to a missing system call related to creating temporary files.
Thus the source code for the CUDA kernels had to be generated on a
different system, copied to TitanDev and the kernels had to be
built explicitly before runtime of the application.
A total of 183 CUDA kernels have been generated in this way.
Since the dynamic linking loader is supported on the Cray XK6 compute nodes
the CUDA kernels could be successfully loaded at program startup.

\subsection{Benchmark Results}

\begin{figure}[t]
\begin{center}
\includegraphics[width=0.75\columnwidth]{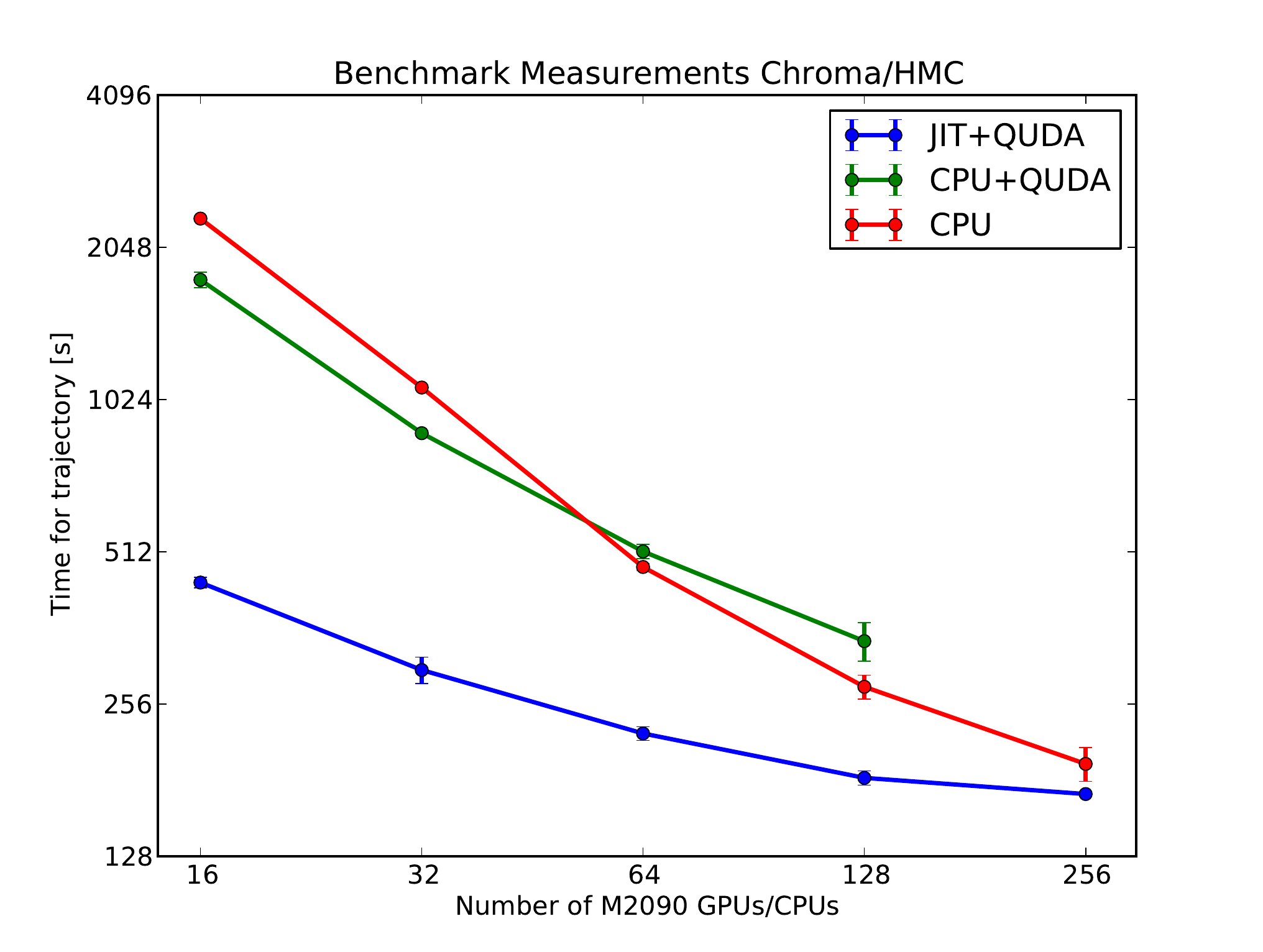}
\end{center}
\vspace*{-3mm}
\caption{\label{fig:hmc}
$N_f=2$ flavors anisotropic Wilson HMC using $32^3\times 96$ lattice
(SP) with Hasenbusch preconditioning}
\end{figure}

 Three different setups were benchmarked:
 \begin{enumerate}
   \item CPU only (QDP++:CPU, Matrix inversion:CPU)
   \item CPU+QUDA (QDP++:CPU, Matrix inversion:GPU) 
   \item JIT+QUDA (QDP++:GPU, Matrix inversion:GPU)
 \end{enumerate}
The time for trajectory was measured for each setup,
Fig. \ref{fig:hmc} shows the result.
For all local volume configurations QDP-JIT together with QUDA
outperforms the other configurations where the CPU is somehow
involved.
The most significant speedup is achieved when the local volume is
reasonably large.
A small speedup is achieved when running on 256 nodes. 
However, arguably the global volume is rather small for this number of
GPUs and at least some strong scaling effects are expected. 

\section{Conclusion}

QDP-JIT provides a 100\% implementation of the QDP++ API for
GPU-enabled parallel systems. It frees the domain expert from tedious
tasks like explicit memory management and CUDA kernel development.
An auto-tuning mechanism ensures always optimal thread grid
configuration.

\section{Outlook}

Adding coalesced memory accesses, interfacing directly with the NVIDIA
driver (thus avoiding the call to NVIDIA NVCC) and overlapping
communications and computation in expressions involving off-node
communications is work in progress.

\section{Acknowledgment}

Data for Fig. \ref{fig:hmc} was generated by B. Joo on the TitanDev
system at the Oak Ridge Leadership Computing Facility (OLCF), 
using data generated with INCITE resources at the OLCF. The 
use of TitanDev at OLCF and GPU nodes from the US National 
Facility for Lattice Gauge Theory  GPU Cluster located at the 
U.S. Department of Energy Jefferson Lab (for kernel pre-generation) is
gratefully acknowledged.

This research was supported by the Research Executive Agency (REA) of
the European Union under Grant Agreement number PITN-GA-2009-238353
(ITN STRONGnet).

\bibliography{lattice12}

\end{document}